\documentclass[aps,prl,showpacs,letterpaper,twocolumn,longbibliography]{revtex4-1}

\usepackage{float}
\usepackage{graphicx} 
\usepackage{array}    
\usepackage{amsmath}   
\usepackage{amssymb}   
\usepackage{comment}

\usepackage{color}  
\usepackage{setspace}
\usepackage{epstopdf}

\newcommand{\unit}[1]{\ensuremath{\, \mathrm{#1}}}

\newcommand{\ket}[1]{| #1 \rangle}
\newcommand{\bra}[1]{\langle #1 |}
\newcommand \bea{\begin{eqnarray}}
\newcommand \eea{\end{eqnarray}}
\newcommand \be{\begin{equation}}
\newcommand \ee{\end{equation}}

\begin{document}	


\title{Doubly magic optical trapping for Cs atom hyperfine clock transitions}
\author{ A. W. Carr and  M. Saffman}
\affiliation{Department of Physics,  1150 University Avenue,
University of Wisconsin,  Madison, Wisconsin 53706
}

 \date{\today}

\begin{abstract}
We analyze doubly magic trapping of Cs hyperfine transitions 
including previously neglected contributions from the 
 ground state hyperpolarizability  and the interaction of the laser light and a static magnetic field. Extensive numerical searches do not reveal any doubly magic trapping conditions for any pair of hyperfine states. However, including the hyperpolarizability reveals light intensity insensitive traps for a wide range of wavelengths at specific intensities.  We then investigate the use of bichromatic trapping light fields.  Deploying a bichromatic scheme, we demonstrate doubly magic red and blue detuned  traps for pairs of states separated by one or two single photon transitions.
\end{abstract}

\maketitle


The international primary standard for time is the Cs atom ground state hyperfine clock transition $|3,0\rangle \leftrightarrow |4,0\rangle$  which is defined to have a frequency of 9192631770 Hz.
Coherent control of alkali atom clock states is of  great interest for precision measurements\cite{Romalis1999} and for encoding  neutral atom qubits for quantum computation experiments\cite{Negretti2011}. Although the clock states 
have excellent coherence properties in a field free environment, fluctuations of optical and magnetic trapping and bias fields lead to differential shifts of the clock state energies causing decoherence.

A large amount of recent work has been devoted to finding  
magic trapping conditions for which variations of external fields do not lead to a differential shift $\delta E$ of the clock state energies.  
For alkali atom hyperfine transitions there are magic conditions 
for hyperfine Zeeman states with $M_F\ne 0$\cite{ChoiCho2007,Flambaum2008,Rosenbusch2009,HKim2013}
which eliminate  sensitivity to fluctuations of the trapping light intensity but are still sensitive to magnetic noise
($d\delta E/d\Omega^2 =0,~ d\delta E/dB\ne 0$ with $\Omega^2$ proportional to the light intensity and $B$ the magnetic field). There are also magic conditions for 
states with $M_F=0$ which are insensitive to light fluctuations at the cost of increased sensitivity to magnetic noise($d\delta E/d\Omega^2 =0,~ d\delta E/dB\gg 0$)  due to the requirement of a relatively large magnetic bias field of several Gauss\cite{Lundblad2010,Dudin2010,Derevianko2010a}. 
In \cite{Derevianko2010b,Chicireanu2011} doubly magic traps were proposed which use $M_F\ne0$ states with elliptically polarized light to cancel the sensitivity to fluctuations of both light intensity and magnetic field strength ($d\delta E/d\Omega^2 =0$ and $d\delta E/dB= 0$). We assume, as is normally done, that there are no fluctuations of the light polarization state. These doubly magic conditions were restricted to certain wavelength ranges and required very precise preparation of the field polarization state.
Doubly magic conditions have also been found for magnetically trapped atoms with microwave frequency dressing fields\cite{Sarkany2014,Kazakov2015}.

\begin{figure}[!t]
\includegraphics[width=8.5cm]{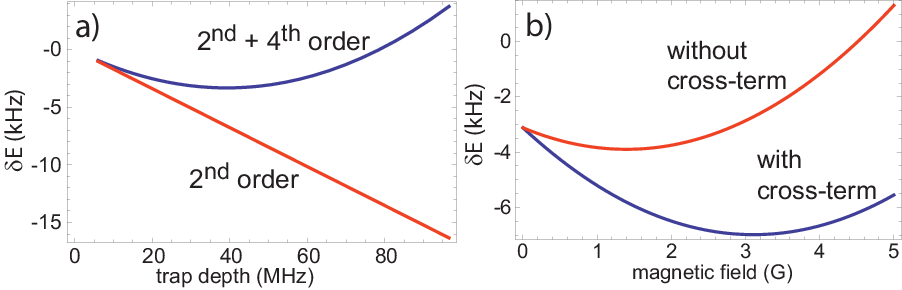}
  \caption{(color online) Differential shift $\delta E$ as a function of trapping light intensity (which is proportional to trap depth) and magnetic field. a)  $\delta E$ at $B=0$ using
2nd order perturbation theory (red) and 4th order (blue) 
for   the  Cs clock states $|3,0\rangle, |4,0\rangle$ at $\lambda=780~\unit{n m}$ with $\sigma_+$ polarized light.  Notice the minimum in $\delta E$ as a function of the light intensity (trap depth) which constitutes a magic operating point. b)   $\delta E$ as a function of the magnetic field with (blue) and without (red) the third order cross-term for $\lambda=780~\unit{n m}$, $\Omega/2\pi =100~\unit{GHz}$, $\sigma_+$ polarization,  and the states $|4,1\rangle, |3,-1\rangle$.}
\label{fig.6p12}
\end{figure}

In this letter we study magic trapping conditions in optical traps while consistently accounting for the hyperpolarizability, which is fourth order in the electric field amplitude, as well as the interaction of the vector polarizability with the static magnetic field. As we illustrate in  Fig. \ref{fig.6p12} and explain below these terms are important for accurate calculations of the differential shift $\delta E$ at typical experimental trap depths. Including these effects, which were not fully accounted for in previous calculations, and after an extensive parameter search we find no parameters for doubly magic trapping in a monochromatic trapping light field.  This suggests that previously reported doubly magic solutions\cite{Derevianko2010b} are not doubly magic. However, we demonstrate insensitivity to differential light shifts with only one frequency and pure circular polarization at specific intensities $\Omega_0^2$ for a wide range of wavelengths ($d\delta E/d\Omega^2|_{\Omega=\Omega_0}=0).$

Although we have not found doubly magic conditions with monochromatic trapping light, a bichromatic approach, following the proposal in \cite{Radnaev2010}, does allow for doubly magic trapping for red or blue detuned optical traps. Our results significantly extend the range of configurations which can be used for doubly magic trapping. In particular, we can  make the the $M_F=0$ clock transition $|4,0\rangle \leftrightarrow |3,0\rangle$ and the two photon transition $|4,1\rangle \leftrightarrow |3,-1\rangle$ doubly magic.  While we provide a viable and flexible approach for achieving extended coherence with trapped alkali atoms we emphasize that  our solutions only give $d\delta E/d\Omega^2|_{\Omega=\Omega_0} =0$ and $d\delta E/dB|_{\Omega=\Omega_0}= 0$ at a specific trapping intensity $\Omega_0^2$. 
Our inclusion of hyperpolarization effects suggests that  conditions for which the derivatives simultaneously vanish independent of the  intensity do not exist. A consequence of this restriction is that   trapped atoms in a thermal motional state  which samples different optical intensities will still experience small time dependent differential shifts.

The calculational approach we follow is to use fine structure energies at zeroth order and treat the hyperfine and Zeeman interactions as perturbations. We use exact diagonalization to consistently include the fourth order hyperpolarizability, third order hyperfine-mediated polarizability, and third order cross-term between the polarizability and the Zeeman interaction. 
An  example of a Hamiltonian between  Cs $6s_{1/2}$ $|F,M_F\rangle$ hyperfine states $\ket{0}=\ket{3,1}$ and $\ket{1}=\ket{4,1}$ is given in Eq. (\ref{eq.Hamiltonian}).
    Though these states are not ideal for magic conditions, for the case of light that is circularly polarized in the plane perpendicular to the magnetic field which defines the quantization axis, their analysis reduces to a simple $2\times2$ matrix which serves to elucidate our procedure for searching for doubly magic traps.   
Our model Hamiltonian for these states is  
\begin{equation}
\mathcal H =\begin{pmatrix}
V_{{\rm hf},0}+V_{00}+\beta_{00}+Z_{00} &  V_{01}+\beta_{01}+Z_{01} \\
 V_{01}+\beta_{01}+Z_{01}  & V_{{\rm hf},1} +V_{11}+\beta_{11}+Z_{11}
\end{pmatrix}.
\label{eq.Hamiltonian}
\end{equation}
Here  $V_{{\rm hf},i}$ is the diagonal hyperfine interaction, $V_{ii},V_{ij}$ arise from the quadratic polarizability due to the trapping light,  
$\beta_{ii}, \beta_{ij}$ are from the third order diagonal and off-diagonal hyperfine-mediated polarizability,
$Z_{ii}$ are the first order diagonal Zeeman shifts, and  $Z_{ij}$ is the magnetic dipole coupling due to the external magnetic field.  For all calculations we include the $6p_{1/2},6p_{3/2},7p_{1/2},7p_{3/2}$ levels, excited $ns_{1/2}$ states up to $14s_{1/2}$ and the counter-rotating terms. For states with different $M_F$ values or elliptically  polarized light the Hamiltonian may have larger dimension and is constructed using the methods detailed in the supplementary material\cite{SMmagic} with analysis based on numerical diagonalization.

\begin{table*}[!ht]
\caption{Intensity magic trap conditions for various wavelengths, light polarizations, pairs of states, and magnetic fields.  Also reported are the first order sensitivities to the magnetic field, and the second order sensitivities to the reduced Rabi frequency.  The polarization is either $\sigma_+$ or $\sigma_-$ and the states column lists $M_{F=4},M_{F=3}$.  Finally, trap depth for blue detuned traps refers to the light shift the atom experiences at the bottom of the trap(see \cite{Piotrowicz2013} for an example of blue traps with nonzero intensity minimum) rather than the actual depth of the trap. The ground state scalar polarizability is negative for the 780 and 820 nm cases so the trap depths are positive, i.e. repulsive potentials. }
\begin{ruledtabular}
\begin{tabular}{*{8}{c}}
$\lambda$ (nm) & $\sigma$ &states & $B$ (G) & $\Omega/2\pi$ (GHz) &  trap depth (MHz)& $\frac{d\delta E}{dB}$ (Hz/G) & $\frac{d^2\delta E}{d\Omega_0^2}$ $(10^{-18}{\rm Hz}^{-1})$ \\ \hline
1038	&+&	0,0&	1.4&	158.0&	-60.0&	-2440&	2.50\\
1038	&+&	0,0&	1.0&	150.0&	-54.3&	-2440&	2.26\\
1038	&+&	0,0&	0&	129.0&	-40.3&	-2440&	1.60\\
1038	&+&	1,-1&	1.4&	164&	-66.5&	-3680&	2.53\\
945	&+&	-1,1&	1.4&	70.8&	-23.3&	-1000&	10.4\\
820	&+&	0,0&	1.4&	70.4&	35.5&	-3680&	22.4\\
780	&+&	0,0&	1.4&	124&	49.2&	-2720&	4.68\\
780	&-&	1,-1&	1.4&	85.6&	24.3&	1760&	2.10\\

\end{tabular}
\end{ruledtabular}
\label{tab.singlefreq}
\end{table*}

\begin{figure}[!t]
\centering
\includegraphics[width=8.5cm]{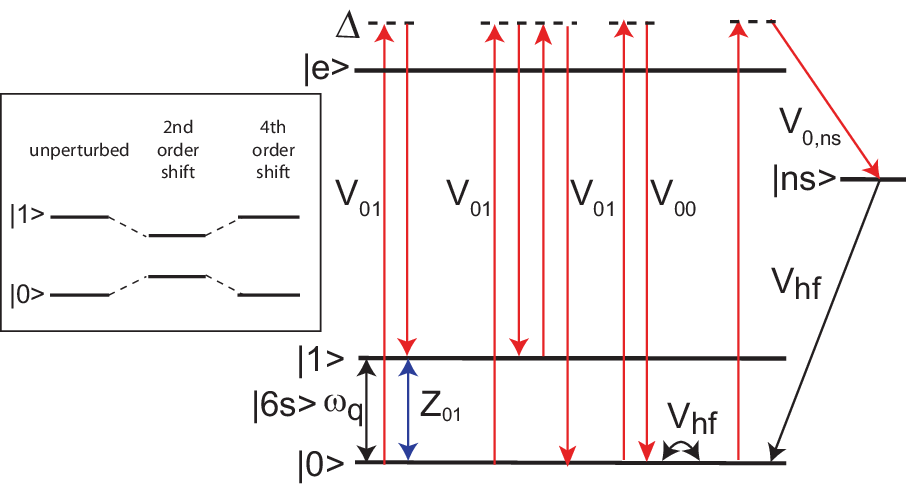}
  \caption{(color online) Schematic representation of four of the terms important to calculating the differential shift where $V_{ij}$ indicates the polarizability between  states $i$ and $j$.  From left to right we have the cross-term of two electric dipole couplings and a magnetic dipole coupling.  This can also happen in reverse order thus the extra factor of two in Eq. (\ref{eq.deltaE}).  Next is the fourth order term with four electric dipole couplings.  This is also the crucial term in the bichromatic scheme except that it then couples to the other ground state that is dressed by a sideband energy very close to the clock frequency.  Third, we have the hyperfine-mediated polarizability term that is diagonal in the ground state hyperfine interaction (HFI). Last is the hyperfine-mediated polarizability term involving HFI-induced mixing of the ground $6s_{1/2}$ state and excited $ns$ states.}
\label{fig.Perturb}
\end{figure}

The eigenvalues of (\ref{eq.Hamiltonian}), replacing the diagonals by $\Delta E_0,\Delta E_1$, are 
$$
\delta E_{1,0}=\frac{\Delta E_0+\Delta E_1 \pm \sqrt{4(V_{01}+Z_{01})^2+(\Delta E_1- \Delta E_0)^2}}{2}
$$
where we ignore $\beta_{01}$ for the moment 
since $V_{01}\gg\beta_{01}$. In the limit where  the hyperfine interaction dominates other perturbations, i.e. $(\Delta E_1- \Delta E_0)^2\gg (V_{01}+Z_{01})^2$, the differential shift relative to the hyperfine splitting $V_{{\rm hf},1}-V_{{\rm hf},0}$ is
\begin{equation}
\delta E=\beta_{11}-\beta_{00}+Z_{11}-Z_{00}+2\frac{Z_{01}^2+2V_{01} Z_{01}+V_{01}^2}{\Delta E_1- \Delta E_0}.
\label{eq.deltaE}
\end{equation}
$V_{00}=V_{11}$ since the electric dipole operator only couples to the electronic quantum numbers which are the same for both states and thus they cancel out of the differential shift.  From here we can make the connection to non-degenerate perturbation theory.  The final term on the right 
includes the second order Zeeman shift $(Z_{01}^2)$, the third order cross-term that is two electric dipole matrix elements and one magnetic dipole coupling matrix element $(2 V_{01} Z_{01} )$ and finally a hyperpolarizability term $(V_{01}^2)$ which is fourth order in the electric dipole operator.  Figure \ref{fig.Perturb} depicts the physical origin of the higher order terms.

 To check our results we calculated the differential light shift per unit of intensity for the case of clock states $M_F=0$ and light linearly polarized perpendicular to the quantization axis as in the experiment of Ref.  \cite{Rosenbusch2009}. In this case the only non-zero off-diagonal term in (\ref{eq.Hamiltonian}) is $Z_{01}$.   For 780 nm light they record $-2.27(40)\times10^{-2}\unit{Hz/W/cm^2}$ compared to our value of $-2.02\times10^{-2}\unit{Hz/W/cm^2}$ and for 532 nm light they measured $-3.51(70)\times10^{-4}\unit{Hz/W/cm^2}$ whereas we calculate $-4.08\times10^{-4}\unit{Hz/W/cm^2}$.  Thus, in both cases we are within experimental error bars and differ by no more than 15\%.

We proceed to write down a system of approximate equations that guide 
the search for doubly magic conditions.   To isolate the dependence on the amplitude of the optical trapping field 
$\mathcal{E}$ and the magnetic field $B$ we make the replacements: 
$\Omega=\frac{{\mathcal E}}{\hbar}\langle 6p_{1/2}||er||6s_{1/2}\rangle$, 
$\beta_{11}-\beta_{00}\rightarrow\beta^{(1)} \Omega^2$,
 $Z_{11}-Z_{00}\rightarrow\mu^{(1)}B$, 
$2\frac{Z_{01}^2}{\Delta E_1- \Delta E_0}\rightarrow\mu^{(2)}B^2$, 
$4\frac{V_{01} Z_{01}}{\Delta E_1- \Delta E_0}\rightarrow\beta^{(2)} \Omega^2B$,
 $2\frac{V_{01}^2}{\Delta E_1- \Delta E_0}\rightarrow\beta^{(4)} \Omega^4$.   Then the differential shift is
\begin{equation}
\delta E=\mu^{(1)}B+\mu^{(2)}B^2+\beta^{(1)}\Omega^2+\beta^{(4)} \Omega^4+\beta^{(2)} \Omega^2B
\label{eq.dE}
\end{equation}
and after taking the appropriate derivatives
\begin{align}
\frac{d\delta E}{dB}&=\mu^{(1)}+2\mu^{(2)}B+\beta^{(2)}\Omega^2 \label{magicB},\\ 
\frac{d\delta E}{d\Omega^2}&=\beta^{(1)}+\beta^{(2)} B+2\beta^{(4)} \Omega^2. \label{magicE}
\end{align}
Doubly magic trapping occurs when both derivatives simultaneously vanish for a set of  parameters $(\Omega, B)$, 
 thus eliminating the first order sensitivity to both electric and magnetic fields.  We emphasize that solving the above equations only yields approximate results since we have suppressed additional small terms arising from the higher order dependence of the coefficients on $\Omega$ and $B$. The accurate results reported in the tables  are found from diagonalizing the ground state Hamiltonian (\ref{eq.Hamiltonian}) and looking for local minima (or maxima) in the differential shift corresponding  to a doubly magic operating point.

It is the implications of the cross-term with 
coefficient $\beta^{(2)}$, and $\beta^{(4)}$ the hyperpolarizability, that drive the novel results in this paper.  The cross-term has been partially included before in \cite{Derevianko2010b} where doubly magic wavelengths were calculated by determining the magic magnetic field with no light interacting with the atoms, i.e. ignoring the last term in Eq. (\ref{magicB}).  However, this neglects the fact that through this term the light intensity also affects the magic magnetic field value, that is, we have a coupled system of equations.  In addition the effects of the fourth order term at the end of Eq. (\ref{magicE}) have not previously been accounted for.  As we will show, this term cannot be neglected and the implication is that there exist not magic wavelengths, but magic intensities for most wavelengths.

The influence of the hyperpolarizability on magic trapping conditions has been shown to be important for optical atomic clocks\cite{Taichenachev2006}, but has   been neglected in previous analyses of microwave clocks.   When field strengths reach 
$10^8~\rm V/m$ the hyperpolarizability cannot be neglected when calculating energy shifts of alkali atom ground states\cite{Manakov1986}.
To put this value in perspective an optical dipole trap for Cs atoms based on a modest
power of  20 mW at $\lambda=1.06~\mu\rm m$ focused to a waist ($1/e^2$ intensity radius) of 
$w=2~\mu\rm m$ gives a field strength $> 10^6~\rm V/m$ and a  trap depth of 15.5 MHz. For detunings large compared to the excited state hyperfine structure we can estimate when the contribution of the fourth order term  to the differential shift becomes as important as that of the hyperfine-mediated polarizability, i.e. the magic intensity point $|\Omega_{\rm 0}|^2$.  We have  $\beta^{(1)}\sim -\omega_q\Omega^2/\Delta^2$ and $\beta^{(4)}\sim\Omega^4/(\omega_q\Delta^2)$, with $\omega_q$ the hyperfine splitting of the two states and $\Delta$ the detuning from the excited state, which suggests the magic intensity scales as 
 $I_{\rm 0}\sim\Omega_{\rm 0}^2\simeq\omega_q^2$.  The implications of this relationship are that the hyperpolarizability becomes important at lower intensities in atomic species with smaller ground state hyperfine splitting, e.g. Rubidium compared to Cesium. Fig. \ref{fig.6p12} illustrates the importance of the fourth order and cross-terms by graphing the differential shift with and without them.

The primary lesson to be drawn from these two terms is that it is inappropriate to talk about magic wavelengths. Even if searching purely for magic conditions in one field the operating laser intensity and magnetic fields are interdependent and the system is now nonlinear in the intensity.  Adjusting the laser intensity then tunes the relative strength of these terms and allows for magic conditions at a more diverse set of wavelengths.

\begin{table*}[!ht]
\caption{Doubly magic conditions for red and blue detuned traps for different wavelengths, pairs of states,  sideband strengths (ratio of sideband field amplitude to carrier field amplitude), sideband frequency,  and the residual sensitivities to field fluctuations.   }
\begin{ruledtabular}
\begin{tabular}{*{9}{c}}
$\lambda$ (nm) & states & SB strength& $\omega_{\rm s}/2\pi$ \rm (GHz)&  B (G) & $\Omega/2\pi$ (GHz)  & trap depth (MHz) & $\frac{d^2\delta E}{dB^2}$ ($\frac{{\rm Hz}}{{\rm G}^2}$) & $\frac{d^2\delta E}{d\Omega^2}$ ($10^{-18} {\rm Hz}^{-1}$)\\ \hline
1038	&0,0	&0.1&	9.15	&1.35	&88.3	&-19	&854	&2.49\\
1038	&0,0	&1	&8	&1.39	&63.6	&-19.5	&854	&4.97\\
1038	&1,-1&	1	&8	&3.63	&80.7	&-32.2	&801	&7.5\\
1038	&-1,1&	0.1	&9	&1.78	&135	&-43.5	&801	&2.59\\
920	&0,0&	0.1	&9.185	&0.364	&17.7	&2.26	&854	&18.5\\
920	&0,0&	0.1	&9.15	&1.54	&27.2	&-5.89	&854	&34.6\\
780	&0,0&	0.1	&9.185&	0.27	&29.9	&2.89	&854	&3.56\\
780	&0,0&	0.1	&9.17&	0.798	&51.4	&8.56	&854	&4.1\\

\end{tabular}
\end{ruledtabular}
\label{tab.doublefreq}
\end{table*}

Unfortunately we were unable to find doubly magic conditions for monochromatic trapping light of any wavelength and polarization for any pair of states.  It is possible that some elliptical polarization we did not check or just a more precise calculation could turn up something useful.  What we can say is that for most wavelengths, magnetic field strengths, pairs of states and polarizations there is a magic intensity where first-order sensitivity to the light intensity is zero.  Table \ref{tab.singlefreq} provides examples of magic intensities for various atomic and field parameters, as well as the residual field sensitivities.  As one can see, despite not being truly doubly magic, the sensitivity to the magnetic field can be quite small for some magic intensity conditions, in particular at 945 nm with the unorthodox $\ket{4,-1}$ and $\ket{3,1}$ states. If one can stabilize the magnetic field to $100~\unit{\mu G}$ then the above states can have differential shifts of about $0.1 ~\unit{Hz}$.

It is worth pointing out some of the trends in  Table \ref{tab.singlefreq}.  Increasing the bias magnetic field increases the trap depth of the magic intensity for the states $0,0$.  Similarly, increasing the detuning increases the trap depth and decreases the residual sensitivity to the trapping light.  We chose $1.4~ \unit{G}$ for the magnetic field since it is very close to the magic magnetic field for the $M_F=0$ states with no trapping light and thus elucidates how important a proper accounting of the aforementioned cross-term is for accurately assessing magic conditions.

In order to achieve true doubly magic traps we adopt the idea of Radnaev et al.\cite{Radnaev2010}.  They envisioned applying another laser that would couple the two ground $6s$ hyperfine levels via a two-photon transition where the  photon fields differ in frequency by very near the hyperfine splitting of the two ground states.  Practically, this could be accomplished by a high frequency phase 
modulator that  adds a frequency shifted sideband at  approximately the hyperfine ground splitting, e.g. $\omega_{\rm s}\sim \omega_{\rm q}=9.192~ \unit{GHz}$ in Cesium.  Thus, we obviate the need for another laser and automatically match the intensity profile of the second light frequency to that of the primary trapping light.

We modify our computational apparatus to incorporate dressed ground states that differ by plus or minus a sideband frequency.  Essentially, the Hamiltonian  matrix grows three fold as for each ground state we add two levels with identical quantum numbers to the original except that the diagonal element has $\pm \omega_s$, the difference between the carrier and sideband frequencies.  The matrix is then populated as before.  We include an explicit example in \cite{SMmagic}. A fourth order term that is roughly $V^s_{ij}V^s_{ij}/(\omega_q-\omega_s)$ arises, where $V^s_{ij}$ uses the field amplitude of the sideband and the carrier, i.e. $\Omega^2\rightarrow\Omega\Omega_s$ with $\Omega_s$ the reduced Rabi frequency of the sideband.  This term is much like the fourth order term introduced for a single frequency in that it couples a ground state to the other ground state and back, but in this case that other state is a dressed state nearly resonant with the initial state due to the sideband.  The denominator depends on the difference of the sideband frequency $\omega_{\rm s}$ and the  clock frequency $\omega_{\rm q}$.  This then gives us two levers with which to adjust this fourth order term's magnitude and sign: the intensity of the sideband and its frequency.  We assume pure $\sigma_+$ polarization for all results regarding the bichromatic scheme.

In Table \ref{tab.doublefreq} we present the results of our search for doubly magic conditions with physically feasible parameters.  As can be seen, doubly magic conditions exist for blue and red detuned traps and multiple pairs of states.  Bringing the sideband frequency closer to the clock frequency lowers the magic magnetic field, the trap depth and the residual sensitivity to the trapping light.  Reducing the sideband intensity reduces the residual light sensitivity for a given magic trap depth and magnetic field.  Lastly, the residual magnetic field sensitivity is entirely determined by the pairs of states chosen and even then is approximately the same for all pairs at the doubly magic point.  Thus, bichromatic trapping light allows one to not only create a doubly magic trap but also to tune the parameters of the trap by adjusting the sideband strength and frequency.

We conclude by reiterating our results for doubly magic trapping conditions.   Our analysis established that the fourth order Stark shift term is vital to differential Stark shift calculations at typical operating conditions and further explored the implications of the interaction of the laser and magnetic fields. Recent experiments have measured the fourth order Stark shift contribution to the differential light shift with Rb atoms and demonstrated the importance of this effect\cite{JYang2016}.    With these tools we demonstrate light insensitive traps for a wide array of wavelengths, pairs of states, and bias magnetic fields.  We then extended our analysis to a bichromatic trap with two optical frequencies separated by an amount similar to the splitting of the hyperfine ground states in Cesium.  In doing so, we discovered doubly magic conditions in red and  blue detuned traps for states separated by only one or two photons.  These results provide a method for obtaining insensitivity to trapping lasers and magnetic field noise that could potentially improve atomic clock and quantum information experiments.  Furthermore, we have relaxed many of the stringent requirements on wavelength, polarization, and states previously reported in the literature, making  magic trapping more accessible to future experiments. We anticipate that similar results will be found for other alkali atom species. 

This work was supported by NSF award PHY-1104531, the
AFOSR Quantum Memories MURI, and the IARPA MQCO
program through ARO contract W911NF-10-1-0347.



%

\pagebreak 
\hspace{1 cm}

\newpage

\newpage

\widetext

\section{Supplementary information for: Doubly magic optical trapping for Cs atom hyperfine clock transitions}

\subsection*{ A. W. Carr and  M. Saffman}

\subsection*{Department of Physics,  1150 University Avenue,
University of Wisconsin,  Madison, Wisconsin 53706
}

\subsection*{\today}

In this supplementary material we provide details of the atomic calculations used to find magic conditions for monochromatic and bichromatic traps.

\section{Light - atom interaction with one frequency of light}

We adopt an effective Hamiltonian approach between the ground states of Cesium.  It is effective by virtue of the fact that no excited states explicitly appear in the Hamiltonian, instead our off-diagonal coupling terms account for the virtual processes through the excited states.  We again write down an example Hamiltonian between $\ket{0}=\ket{F=3,M_F=1}$ and $\ket{1}=\ket{F=4,M_F=1}$ before delving into the computational specifics.
$$
\mathcal H =\begin{pmatrix}
V_{{\rm hf},0}+V_{00}+\beta_{00}+Z_{00} &  V_{01}+\beta_{01}+Z_{01} \\
 V_{01}+\beta_{01}+Z_{01}  & V_{{\rm hf},1} +V_{11}+\beta_{11}+Z_{11}
\end{pmatrix}
$$
We have left the hyperfine interaction (HFI) as a perturbation term and thus all of the polarizability terms use fine structure energy denominators. For the second order polarizability we use the notation of \cite{LeKien2013}

\begin{eqnarray}
V_{ij}&=&\frac{\Omega^2}{4}\sum_{\substack{K=0,1,2\\q=-K,..,K}}\alpha^{(K)}_{nJ}\{u^*\otimes u\}_{Kq}(-1)^{J+I+q-M_i+3F_i-M_j}\sqrt{2F_i+1}
C_{F_iM_iKq}^{F_jM_j}
\begin{Bmatrix}
F_i & K & F_j\\
J & I & J
\end{Bmatrix}
\end{eqnarray}
with the following further definitions for the reduced polarizability and polarization  tensor
\begin{eqnarray}
\alpha^{(K)}_{nJ}&=&(-1)^{K+J}\sqrt{2K+1}
\sum_{n'J'}(-1)^{J'}
\begin{Bmatrix}
1 & K & 1\\
J & J' & J
\end{Bmatrix}
\left|\frac{\langle n'J'||r||nJ\rangle}{\langle 6p_{1/2}||r||6s_{1/2}\rangle}\right|^2
\left(\frac{1}{\omega-\omega_{n'J'nJ}}+\frac{(-1)^{(K+1)}}{\omega+\omega_{n'J'nJ}}\right),
\end{eqnarray}

\begin{eqnarray}
\{\mathbf{u}^*\otimes\mathbf{u}\}_{Kq}&=&\sum_{m,m'=0,\pm1} u^*_{m} u_{m'}C_{1,m,1,m'}^{K,q}\, ,
\end{eqnarray}
with $K=0,1,2$ a tensor rank corresponding to the  scalar, vector and tensor polarizabilities, $q$ a spherical tensor component, $\mathbf{u}$ a polarization vector written in the spherical tensor basis, $I$ the nuclear spin and $J,J'$ the total spin and orbital angular momentum of the ground and excited state, respectively. $\Omega=\frac{{\mathcal E}}{\hbar}\langle 6p_{1/2}||er||6s_{1/2}\rangle$ is the reduced Rabi frequency with ${\mathcal E} = (2 I_0/\epsilon_0 c)^{1/2}$, $I_0$ is the optical intensity and $e$ is the elementary electric charge. As stated before $\omega_{n'J'nJ}$ is the energy difference between the ground state and an excited state without any hyperfine splitting, i.e. just fine structure energies. We write all of our results in terms of the reduced Rabi frequency, $ \Omega$.  Finally, we only included four excited $np$ states in our calculations $6p_{1/2}, 6p_{3/2}, 7p_{1/2}, 7p_{3/2}$. The reduced matrix elements from the $6s_{1/2}$ ground state to these states are given in Sec. \ref{sec.numbers}.  The static polarizability is dominated by the $6p$ terms, on the order of 96\% according to \cite{Iskrenova-Tchoukova2007}, and the dynamic polarizabilities near these transitions are even more dominated by $6p$ since the energy denominators are even smaller for these states by roughly a factor of $\Delta/\omega_{6s6p}$ with $\Delta$ the detuning from $6p$ while leaving relatively unchanged those of other states.  The only other notable contribution is from $7p$, being the next allowed dipole transition, but as can be seen the reduced radial matrix element is an order of magnitude smaller along with carrying a much larger energy denominator for the cases we study.

Next is the hyperfine-mediated polarizability, $\beta$, that consists of two electric dipole couplings and a hyperfine interaction (HFI) coupling. This can be split into 4 broad categories based on the nature of the HFI matrix element: diagonal in the ground state, off-diagonal mixing of the ground state with other $ns$ and $nd$ levels, diagonal in the excited $np$ levels, and off-diagonal mixing of excited $np$ levels.  We only account for those terms with a HFI diagonal in the ground states and those that mix the $6s$ ground state with excited $ns$ states, as including other terms results in a smaller than 1\% correction.  More detailed  discussion of the hyperfine-mediated polarizability can be found in   \cite{Rosenbusch2009SM},  \cite{Ulzega2006,*Ulzega2007,*Hofer2008}.

The term with the HFI matrix diagonal in the ground state term can be written as 
$$
\beta'_{ii}= -V_{{\rm hf},i}V_{ii}',
$$
where 
$V_{ii}' = V_{ii} $ with the replacement $(\omega\pm\omega_{n'J'nJ})\rightarrow({\omega\pm\omega_{n'J'nJ}})^2$, and 
 $V_{{\rm hf},i}=\langle 6s_{1/2}F_iM_i|V_{\rm hf}|6s_{1/2}F_iM_i\rangle$ the diagonal matrix element of the standard magnetic dipole 
hyperfine operator\cite{Arimondo1977} $V_{\rm hf}$ in the ground states.

The other term involving hyperfine mixing of the $s$ ground state with excited $ns$ states is a bit trickier.  We approximate the hyperfine constant characterizing the mixing as \cite{Derevianko1999}
$$
A_{6sns}=\sqrt{A_{6s}A_{ns}}
$$
which is verified empirically to work at the 1\% level.  The matrix element is related to this constant in the usual way
$$
V_{\rm hf,6sns}=\bra{6s_{1/2}FM}V_{\rm hf}\ket{ns_{1/2}FM}=A_{6sns}\mathbf{I}\cdot \mathbf{J}
$$
and the operator appearing in the Hamiltonian  is
$$
\beta''_{ij}=V_{\rm hf,6sns}V_{ij}''
$$
with  $V_{ii}'' = V_{ii} $ where in one reduced matrix element the ground state is replaced by an excited $s$ state,$ \bra{6p_{1/2}}|r|\ket{6s_{1/2}}\rightarrow \bra{6p_{1/2}}|r|\ket{ns_{1/2}}$, and the energy denominator is now $(\omega\pm\omega_{n'J'nJ})\rightarrow(\omega\pm\omega_{n'J'nJ})(\omega_{6s_{1/2},ns_{1/2}})$ and we  include a sum over excited $ns$ states.  For this calculation we only include mixing up to $14s_{1/2}$ and use reduced electric dipole matrix elements calculated from a WKB approximation for elements involving $8s_{1/2}$ or higher.  It is vital to note that the sign of the reduced matrix elements is now crucial since they are no longer squared in the polarizability, $V''$.  

Combining the above contributions the hyperfine-mediated polarizability is 
$$
\beta_{ij}=\beta'_{ij}\delta_{i,j}+\beta''_{ij}
$$
where the first term is diagonal and the second also contributes an off-diagonal vector component.  We keep this vector component in our calculations though it is dwarfed by the second order vector polarizability by a factor that is approximately the hyperfine mixing matrix element divided by the energy difference between the ground state and an excited $s$ state or $\simeq10^{-5}$.

Two sources contribute the majority of the error in our calculation.  First, the accuracy of our WKB approximation of the reduced dipole matrix elements connecting $6p$ and $7p$ to excited $ns$ levels is unknown.  Secondly, $\beta''_{ij}$ is known to converge slowly as a function of the included excited states  and thus our truncation at $14s_{1/2}$ will introduce nontrivial inaccuracies.  Comparison with recent experimental determinations of the DC Stark coefficient, $k_s$, yields an error of approximately 5\%.  A more pertinent test for our calculations is to compare against the experimental data in \cite{Rosenbusch2009SM} for the differential light shift per unit of intensity.  For 780 nm light they record $-2.27(40)\times10^{-2}\unit{Hz/W/cm^2}$ compared to our value of $-2.02\times10^{-2}\unit{Hz/W/cm^2}$ and for 532 nm light they measured $-3.51(70)\times10^{-4}\unit{Hz/W/cm^2}$ whereas we calculate $-4.08\times10^{-4}\unit{Hz/W/cm^2}$.  Thus, in both cases we are within experimental error bars and differ by no more than 15\%.

Finally there are Zeeman terms 
\begin{eqnarray}
Z_{ij}&=&\delta_{M_iM_j}\frac{\mu_B (-1)^{F_i+I+J+1}\sqrt{2F_i+1}}{\hbar}\nonumber\\
&\times&
\left(g_J \sqrt{J(J+1)(2J+1)}
\left\{\begin{matrix}J&I&F_i\\ F_j &1&J \end{matrix}\right\}
+g_I (-1)^{F_j-F_i} \sqrt{I(I+1)(2I+1)}
\left\{\begin{matrix}I&J&F_i\\ F_j &1&I \end{matrix}\right\}
\right)
C_{F_iM_i10}^{F_jM_j}.\nonumber
\end{eqnarray}
Here we have specialized to the case of a static magnetic field polarized along $\hat z$.
The diagonal coeffiients $Z_{ii}$  give the usual expression for the hyperfine Zeeman shifts 
$$
Z_{ii}=\frac{M_i\mu_B g_{F_i} B}{\hbar}.
$$
For the Cs ground state $g_J\simeq 2$, $g_{F=3}\simeq -1/4,\,  g_{F=4}\simeq 1/4$, and 
the nuclear Land\'e factor is  $g_I=-0.00039885395 ~\mu_B = -0.73235679 ~\mu_N $ with $\mu_B$ the Bohr magneton and $\mu_N$ the nuclear magneton. 

\section{Bichromatic light - atom interaction}

\begin{figure}[!t]
\centering
\includegraphics[width=8.5cm]{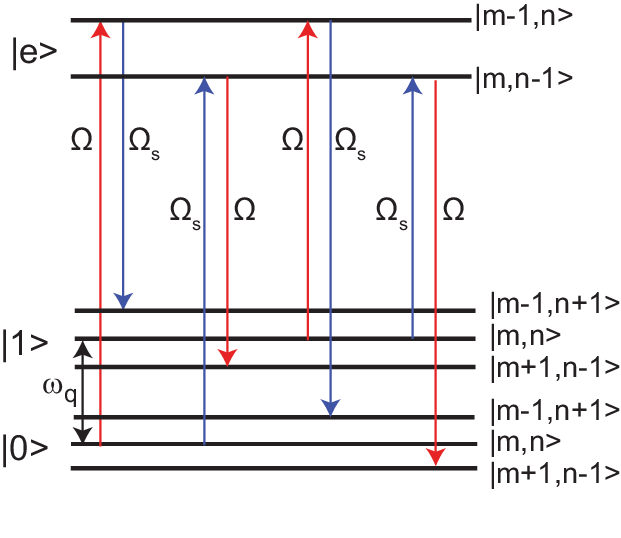}
  \caption{The coupling for the bichromatic scheme with $\Omega$ representing coupling via the carrier laser and $\Omega_s$ that of the sideband.  Four potential terms can contribute, but only two are important at any one time depending on sign of the sideband frequency $\omega_s$.  When $\omega_s\simeq\pm\omega_q$ two of the dressed states are becoming nearly degenerate with a dressed state of the other atomic ground state.  This gives rise to a fourth order term that is tunable by the strength of the sideband and the sideband frequency.}
\label{fig.twocolor}
\end{figure}

We now extend the above Hamiltonian to the two frequency case where we dress the ground states with sideband fields.  The basis is $\ket{0,m+1,n-1},\ket{0,m,n},\ket{0,m-1,n+1},\ket{1,m+1,n-1},\ket{1,m,n},\ket{1,m-1,n+1}$ where $m$ and $n$ are the photon numbers of the two fields with energy $\omega_m$ and $\omega_n$, respectively, and $\omega_s=\omega_n-\omega_m$.  The matrix dimension is then threefold larger than the monochromatic case as each ground state has two additional dressed states that differ in energy by $\pm\omega_s$ from the original state.  In all other respects they are identical and the above definitions for matrix elements are still valid except that we must keep track of when the sideband amplitude is used.  To this end we introduce $V^s$ when the sideband is one of the field amplitudes in the polarizability, $\Omega^2\rightarrow\Omega\Omega_s$, and assume that when $V_{ij}$ appears that one computes it for the carrier amplitude and the sideband amplitude, $\Omega^2\rightarrow\Omega_s^2$ where $\Omega_s$, is the reduced Rabi frequency for the sideband.  Finally, for brevity we make the following substitution $\Delta E_i=V_{{\rm hf},i}+V_{ii}+\beta_{ii}+Z_{ii}$ and omit the vector contribution of the hyperfine-mediated polarizability, though again it is included in our full calculation. We write down a $6\times 6$ example Hamiltonian for the ground states in the above basis with non-elliptical polarization for both fields.

$$
\mathcal H =
\begin{pmatrix}
\Delta E_0-\omega_s & V^s_{00} & 0 & Z_{01} + V_{01} & Z_{01} + V^s_{01} & 0 \\
 V^s_{00} & \Delta E_0 &  V^s_{00} & Z_{01} + V^s_{01} & Z_{01} + V_{01}&Z_{01} + V^s_{01}\\
0 & V^s_{00} & \Delta E_0+\omega_s & 0 & Z_{01} + V^s_{01} & Z_{01} + V_{01}\\
Z_{01} + V_{01} &  Z_{01} + V^s_{01} & 0 & \Delta E_1-\omega_s & V^s_{11} & 0 \\
Z_{01} + V^s_{01} &  Z_{01} + V_{01} & Z_{01} + V^s_{01} &  V^s_{11}&\Delta E_1& V^s_{11}\\
0 & Z_{01} + V^s_{01}& Z_{01} + V_{01} & 0 & V^s_{11} & \Delta E_1+\omega_s
\end{pmatrix}
$$
At this point one can diagonalize and find the eigenvalues of the states $\ket{0,m,n}$ and $\ket{1,m,n}$ to acquire the shifts of the atomic state due to the magnetic and A.C. electric fields.

\section{Numerical values}
\label{sec.numbers}

In this section we list the hyperfine constants and radial reduced matrix elements used for calculations. 
Table I gives hyperfine constants and Tables II-V give radial matrix elements. Where possible we have used numerical values from \cite{Iskrenova-Tchoukova2007}. For other transitions we have used the values found from WKB calculations following the theory of Ref.  \cite{Kaulakys1995}. As a check on the WKB calculations we compare them in the tables to the published values in \cite{Iskrenova-Tchoukova2007} and to Coulomb wave function (Cwf) calculations. The WKB and Cwf calculations use effective quantum defects extracted from the energy values listed in \cite{Kramida2013,Sansonetti2009}. We estimate the error 
of the WKB matrix elements to be less than 10\% on the basis of comparison to values from \cite{Iskrenova-Tchoukova2007}.
 The Cwf and WKB calculations give a value for $R_{nl_{j}}^{nl'_{j'}}$ which is converted to a reduced matrix element using 
$$
\langle n'l'sj' ||r||nlsj \rangle=(-1)^{1+s+j+(1+l+l')/2}\sqrt{(2j+1)(2j'+1){\rm max}(l,l')} 
\left\{\begin{matrix} 
l& s &j\\
j'&1&l'
\end{matrix}\right\}
R_{nl_j}^{n'l'_{j'}}.
$$
 When calculating the third order hf-Stark polarizability the sign of the reduced matrix elements is significant. Note that $\bra{n'j'}|\hat\bf T|\ket{nj}=(-1)^{j-j'}\bra{nj}|\hat\bf T|\ket{n'j'}.$ Thus 
$\bra{n'p_{1/2}}|\hat\bf T|\ket{ns_{1/2}}=\bra{ns_{1/2}}|\hat\bf T|\ket{n'p_{1/2}}$ but $\bra{n'p_{3/2}}|\hat\bf T|\ket{ns_{1/2}}=-\bra{ns_{1/2}}|\hat\bf T|\ket{n'p_{3/2}}$. There is inconsistency in the literature as regards signs of published reduced matrix elements. We have taken the signs from our Cwf or WKB calculations which agree with each other.

This work was supported by NSF award PHY-1104531, the
AFOSR Quantum Memories MURI, and the IARPA MQCO
program through ARO contract W911NF-10-1-0347.

\newpage

\begin{table}[!t]
\setlength{\extrarowheight}{3pt}
\centering
\begin{tabular}{|c|cc|}
\hline
state & $A$ (MHz) & B (MHz) \\
\hline
$6s_{1/2}$ &2298.1579425&\\
$7s_{1/2}$&546.3 &\\
$8s_{1/2}$&218.9 &\\
$9s_{1/2}$&110.1 &\\
$10s_{1/2}$&63.2 &\\
$11s_{1/2}$ &39.4&\\
$12s_{1/2}$ &26.31&\\
$13s_{1/2}$ &18.40&\\
$14s_{1/2}$&13.41&\\
$15s_{1/2}$&$10.1^c$&\\
\hline
$6p_{1/2}$ &$291.9309^c$&\\
$7p_{1/2}$ &94.35&\\
$8p_{1/2}$ &42.97&\\
$9p_{1/2}$ &23.19&\\
\hline
$6p_{3/2}$ &$50.28825^c$&$-0.4940^c$\\
$7p_{3/2}$& 16.605&-0.15\\
$8p_{3/2}$ &7.58&-0.14\\
$9p_{3/2}$ &4.123&-0.051\\
\hline
$5d_{3/2}$ &$48.78^a$&$0.1^a$\\
$6d_{3/2}$ &$16.34^c$ &$-0.1^c$\\
$7d_{3/2}$ &$ 7.4^c$&\\
$8d_{3/2}$ & $3.95^c$&\\
\hline
$5d_{5/2}$ &$-21.24^a$&$0.2^a$\\
$6d_{5/2}$ & $-4.69^b$ & $0.18^b$\\
$7d_{5/2}$ &-1.7&\\
$8d_{5/2}$ & -0.85&\\
\hline
\end{tabular}
\caption{Some hyperfine constants of Cs. Values from \cite{Arimondo1977}  except for a) Ref. \cite{Yei1998}, 
 b) Ref. \cite{Georgiades1994}, c) Ref.  \cite{Sansonetti2009}}.
\label{tab.CshfA}
\end{table}

\begin{table}[!t]
\setlength{\extrarowheight}{3pt}
\centering
\begin{tabular}{|ccc|ccc|}
\hline
final & transition & theory && Cwf& WKB \\
state &wavelength $(\mu\rm m)$ & $\langle nl_{j}||r||6p_{1/2}\rangle^a$  &$r_{\rm min}$ &$\langle nl_{j}||r||6p_{1/2}\rangle$  &$\langle nl_{j}||r||6p_{1/2}\rangle$  \\
\hline
$6s_{1/2}$ &-0.8946 &$4.489$&0.&4.501&4.482\\
$7s_{1/2}$ &1.359& $-4.236$&0.&-3.921&-3.970\\
$8s_{1/2}$ &0.7611& &0.&-0.9379&-0.9650\\
$9s_{1/2}$ &0.6356& &0.&-0.4956&-0.5159\\
$10s_{1/2}$ &0.5840& &0.&-0.3270&-0.3434\\
$11s_{1/2}$ &0.5570& &0.&-0.2396&-0.2533\\
$12s_{1/2}$ &0.5408& &0.&-0.1867&-0.1983\\
$13s_{1/2}$ &0.5303& &0.&-0.1513&-0.1614\\
$14s_{1/2}$ &0.5230&  &0.&-0.1262&-0.1351\\
$15s_{1/2}$ &0.5178&  &0.&-0.1075&-0.1155\\
\hline
$5d_{3/2}$ &3.011&7.06&0.7&7.317&7.056\\
$6d_{3/2}$  &0.8764&-4.15&0.7&-4.002&-4.228\\
$7d_{3/2}$  &0.6725&&0.7&-1.976&-2.070\\
$8d_{3/2}$  &0.6012&&0.7&-1.256&-1.308\\
\hline
\end{tabular}
\caption{Reduced matrix elements of coupled states $\ket{nlsj}$ in the $j$ basis $\langle n'l'sj' ||r||6p_{1/2} \rangle$  for the Cs $6p_{1/2}$ state and transition vacuum wavelengths. 
Cwf are values calculated using Coulomb wavefunctions with experimental values for quantum defects, and $r_{\rm min}$ a small $r$ cutoff to avoid divergence. WKB are values calculated using the theory of \cite{Kaulakys1995}.  Matrix elements are given in atomic units.  a)From  \cite{Iskrenova-Tchoukova2007} with the sign from Cwf calculation. }
\label{tab.6p12rme}
\end{table}

\begin{table}[!t]
\setlength{\extrarowheight}{3pt}
\centering
\begin{tabular}{|ccc|ccc|}
\hline
final & transition & theory && Cwf& WKB \\
state &wavelength $(\mu\rm m)$ & $\langle nl_{j}||r||6p_{3/2}\rangle^a $&$r_{\rm min}$ &$\langle nl_{j}||r||6p_{3/2}\rangle$  &$\langle nl_{j}||r||6p_{3/2}\rangle$  \\
\hline
$6s_{1/2}$ &-0.8523 &$6.324$&0.&6.309&6.2856\\
$7s_{1/2}$ &1.470& $-6.473$&0.&-6.026&-6.0946\\
$8s_{1/2}$ &0.7946&&0.&-1.349&-1.3847\\
$9s_{1/2}$ &0.6588& &0.&-0.7049&-0.7318\\
$10s_{1/2}$ &0.6036& &0&-0.4632&-0.4848\\
$11s_{1/2}$ &0.5747&&0.&-0.3386&-0.3568\\
$12s_{1/2}$ &0.5575&&0.&-0.2634&-0.2789\\
$13s_{1/2}$ &0.5463&&0.&-0.2134&-0.2269\\
$14s_{1/2}$ &0.5386& &0.&-0.1779&-0.1898\\
$15s_{1/2}$ &0.5331& &0.&-0.1515&-0.1621\\
\hline
$5d_{3/2}$ &3.614&$-3.19$&0.7&-3.259&-3.167\\
$5d_{5/2}$ &3.491 &$9.66$&0.7&9.871&9.594\\
$6d_{3/2}$ &0.9211 &$2.05$&0.7&2.003&2.0923\\
$6d_{5/2}$ &0.9175 &$-6.01$&0.7&-5.860&-6.129\\
$7d_{3/2}$ &0.6985& &0.7&0.9524&0.9857\\
$7d_{5/2}$ &0.6975& &0.7&-2.816&-2.919\\
$8d_{3/2}$ &0.6220& &0.7&0.5973&0.6142\\
$8d_{5/2}$ &0.6215& &0.7&-1.773&-1.827\\
\hline
\end{tabular}
\caption{Reduced matrix elements of coupled states $\ket{nlsj}$ in the $j$ basis $\langle n'l'sj' ||r||6p_{3/2} \rangle$  for the Cs $6p_{3/2}$ state and transition vacuum wavelengths. 
Cwf are values calculated using Coulomb wavefunctions with experimental values for quantum defects, and $r_{\rm min}$ a small $r$ cutoff to avoid divergence. WKB are values calculated using the theory of \cite{Kaulakys1995}.  Matrix elements are given in atomic units.  a)From  \cite{Iskrenova-Tchoukova2007} with the sign from Cwf calculation. }
\label{tab.6p32rme}
\end{table}

\begin{table}[!t]
\setlength{\extrarowheight}{3pt}
\centering
\begin{tabular}{|ccc|ccc|}
\hline
final & transition & theory && Cwf& WKB \\
state &wavelength $(\mu\rm m)$ & $\langle nl_{j}||r||7p_{1/2}\rangle^a$  &$r_{\rm min}$ &$\langle nl_{j}||r||7p_{1/2}\rangle$  &$\langle nl_{j}||r||7p_{1/2}\rangle$  \\
\hline
$6s_{1/2}$ &-0.4594 &$0.276$&0.&0.465& 0.4175\\
$7s_{1/2}$ &-3.096& $10.308$&0.&10.18&10.18\\
$8s_{1/2}$ &3.919&-9.313 &0.&-9.115&-9.154\\
$9s_{1/2}$ &1.943& &0.&-1.921&-1.940\\
$10s_{1/2}$ &1.530& &0.&-0.9761&-0.9909\\
$11s_{1/2}$ &1.358& &0.&-0.6338&-0.6460\\
$12s_{1/2}$ &1.265& &0.&-0.4612&-0.4716\\
$13s_{1/2}$ &1.209& &0.&-0.3583&-0.3674\\
$14s_{1/2}$ &1.172&  &0.&-0.2904&-0.2985\\
$15s_{1/2}$ &1.146&  &0.&-0.2424&-0.2497\\
\hline
$5d_{3/2}$ &-1.376&&0.7&-1.897&-2.091\\
$6d_{3/2}$  &12.14&18.0&0.7&17.92&17.86\\
$7d_{3/2}$  &2.335&6.56&0.7&-6.402&-6.519\\
$8d_{3/2}$  &1.654&&0.7&-3.124&-3.193\\
\hline
\end{tabular}
\caption{Reduced matrix elements of coupled states $\ket{nlsj}$ in the $j$ basis $\langle n'l'sj' ||r||7p_{1/2} \rangle$  for the Cs $7p_{1/2}$ state and transition vacuum wavelengths. 
Cwf are values calculated using Coulomb wavefunctions with experimental values for quantum defects, and $r_{\rm min}$ a small $r$ cutoff to avoid divergence. WKB are values calculated using the theory of \cite{Kaulakys1995}.  Matrix elements are given in atomic units.   a)From  \cite{Iskrenova-Tchoukova2007} with the sign from Cwf calculation.  }
\label{tab.7p12rme}
\end{table}

\begin{table}[!t]
\setlength{\extrarowheight}{3pt}
\centering
\begin{tabular}{|ccc|ccc|}
\hline
final & transition & theory && Cwf& WKB \\
state &wavelength $(\mu\rm m)$ & $\langle nl_{j}||r||7p_{3/2}\rangle^a$  &$r_{\rm min}$ &$\langle nl_{j}||r||7p_{3/2}\rangle$  &$\langle nl_{j}||r||7p_{3/2}\rangle$  \\
\hline
$6s_{1/2}$ &-0.4557 &$0.586$&0.&0.8158& 0.7534\\
$7s_{1/2}$ &-2.932   &14.320 $ $&0.&14.13&14.13\\
$8s_{1/2}$ &4.218&  &0.&-13.78&-13.84\\
$9s_{1/2}$ &2.014& &0.&-2.681&-2.705\\
$10s_{1/2}$ &1.574& &0.&-1.343&-1.361\\
$11s_{1/2}$ &1.392& &0.&-0.8672&-0.8825\\
$12s_{1/2}$ &1.295& &0.&-0.6294&-0.6424\\
$13s_{1/2}$ &1.236& &0.&-0.4882&-0.4996\\
$14s_{1/2}$ &1.197&  &0.&-0.3953&-0.4054\\
$15s_{1/2}$ &1.170&  &0.&-0.3298&-0.3389\\
\hline
$5d_{3/2}$ &-1.343&&0.7&0.7573&0.8310\\
$5d_{5/2}$ &-1.361&&0.7&-2.369&-2.591\\
$6d_{3/2}$  &15.57&-8.07&0.7&-8.029&-8.011\\
$6d_{5/2}$  &14.59&24.4&0.7&24.22&24.16\\
$7d_{3/2}$  &2.438&3.32&0.7&3.253&3.300\\
$7d_{5/2}$  &2.426&-9.64&0.7&-9.459&-9.604\\
$8d_{3/2}$  &1.705&&0.7&1.520&1.546\\
$8d_{5/2}$  &1.702&&0.7&-4.473&-4.554\\
\hline
\end{tabular}
\caption{Reduced matrix elements of coupled states $\ket{nlsj}$ in the $j$ basis $\langle n'l'sj' ||r||7p_{3/2} \rangle$  for the Cs $7p_{3/2}$ state and transition vacuum wavelengths. 
Cwf are values calculated using Coulomb wavefunctions with experimental values for quantum defects, and $r_{\rm min}$ a small $r$ cutoff to avoid divergence. WKB are values calculated using the theory of \cite{Kaulakys1995}.  Matrix elements are given in atomic units.  a)From  \cite{Iskrenova-Tchoukova2007} with the sign from Cwf calculation.  }
\label{tab.7p32rme}
\end{table}

\end{document}